\documentclass[prd,floatfix,preprintnumbers]{revtex4}
\usepackage{amsmath}
\usepackage{graphicx,epsfig}
\usepackage{epstopdf}
\usepackage{color}
\begin{document}
\title{The Quadratic Approximation for Quintessence with Arbitrary
Initial Conditions}
\author{Jeffrey R. Swaney}
\affiliation{Department of Physics and Astronomy, Vanderbilt University,
Nashville, TN  ~~37235}
\affiliation{Department of Physics and Astronomy, University of California,
Irvine, CA ~~92697}  
\author {Robert J. Scherrer}
\affiliation{Department of Physics and Astronomy, Vanderbilt University,
Nashville, TN  ~~37235}

\begin{abstract}
We examine quintessence models for dark energy in which the scalar field, $\phi$,
evolves near the vicinity of a local maximum or minimum in the potential $V(\phi)$, so that $V(\phi)$ be approximated by a
quadratic function of $\phi$ with no linear term.  We generalize previous studies of this type by allowing the initial value of $d \phi/dt$ to
be nonzero.  We derive an analytic approximation for $w(a)$ and show that it is in excellent agreement with numerical simulations
for a variety of scalar field potentials having local minima or maxima.  We derive an upper bound
on the present-day value of $w$ as a function of the other model parameters and present
representative limits on these models from observational data.  This work represents a final generalization of previous
studies using linear or quadratic approximations for $V(\phi)$.
\end{abstract}

\maketitle

\section{Introduction}

Cosmological data \cite{Knop,Riess,union08,perivol,hicken,Hinshaw,Ade}
indicate that roughly
70\% of the energy density in the
universe is in the form of a negative-pressure component,
called dark energy, with roughly 30\% in the form of nonrelativistic matter (including both baryons
and dark matter).
The dark energy component can be parametrized by its equation of state parameter, $w$,
defined as the ratio of the dark energy pressure to its density:
\begin{equation}
\label{w}
w=p/\rho,
\end{equation}
where a cosmological constant, $\Lambda$, corresponds to the case $w = -1$ and $\rho = constant$.
While a model with a cosmological constant and cold dark matter ($\Lambda$CDM) is consistent
with current observations,
there are many realistic models of the Universe that have a dynamical equation
of state.
For example, one can consider quintessence models, with a time-dependent scalar field, $\phi$,
having potential $V(\phi)$
\cite{Wetterich,RatraPeebles,CaldwellDaveSteinhardt,LiddleScherrer,SteinhardtWangZlatev}.
(See Ref. \cite{Copeland} for a review).

In practice, an enormous number of quintessence models that provide an
acceptable fit to the data can be (and have been) developed \cite{Copeland}.  The goal of this paper is to further develop
ideas presented in Refs. \cite{ScherrerSen1,ScherrerDutta1,ds1,ds3}, which systematized and classified
the evolution of quintessence models.  (See also Ref. \cite{Gong}, which takes a similar,
but somewhat different approach). The main idea of those papers was to determine whether
a wide class of quintessence models could yield a single form for the evolution for $w(a)$, or a set
of $w(a)$ trajectories depending on just a few free parameters.
The starting point for Refs. \cite{ScherrerSen1,ScherrerDutta1,ds1,ds3}
is the observational fact that $w$ is very close to $-1$ at present.
Requiring $w \approx -1$ at all earlier times allows for an enormous simplication of the equations
governing the evolution of the quintessence field.

One way to achieve a value of $w$ close to $-1$ is
for $\phi$ to be located in a very flat portion of the potential, so that
\begin{equation}
\label{flat}
\left(\frac{V^\prime}{V}\right)^2 \ll 1,
\end{equation}
where $V^\prime \equiv dV/d\phi$.
Ref. \cite{ScherrerSen1} investigated models
in which equation (\ref{flat}) is satisfied and $V^\prime/V$ is roughly constant,
which will be the case as long as
\begin{equation}
\label{linear}
\left|\frac{V^{\prime \prime}}{V} \right| \ll 1,
\end{equation}
where $V^{\prime \prime} \equiv d^2 V/d\phi^2$.
Ref. \cite{ScherrerSen1} imposed the additional constraint that the field
be nearly static at some initial time, so that $\dot \phi_i = 0$, where the dot will denote a
time derivative throughout, and the subscript $i$ will refer to an arbitrary fixed initial time.  In
the terminology of Caldwell and Linder \cite{CL}, these are ``thawing" models.
When these conditions are satisfied, $w(a)$ takes on a single functional form determined only
by the present-day values of $w$ and $\Omega_{\phi}$ (the fraction of the total density
contributed by the scalar field), but otherwise containing no free parameters.  In Ref.
\cite{ScherrerDutta1}, the results of Ref. \cite{ScherrerSen1} were extended to the case where
$\dot \phi_i \ne 0$.  This yields an expression for $w(a)$ that depends on
$\Omega_{\phi 0}$ and $w_0$ (where the $0$ subscript will refer to present-day values throughout),
but with an additional free parameter determined by $\dot \phi_i$ or, equivalently, by $w_i$.

While Eqs. (\ref{flat}) and (\ref{linear}) are sufficient to ensure that
$w\approx -1$, Eq. (\ref{linear}) is not necessary.  Hence, Refs. \cite{ds1,ds3}
considered potentials in which Eq. (\ref{flat}) is satisfied, but Eq. (\ref{linear})
is not.  These correspond to evolution close to a local
maximum \cite {ds1} or minimum \cite{ds3} in the potential, for which $V(\phi)$ can
be well-approximated as
\begin{equation}
\label{quadratic}
V(\phi) \approx V(\phi_*) + \frac{1}{2}V''(\phi_*) (\phi-\phi_*)^2,
\end{equation}
where the maximum or minimum in the potential is located at $\phi = \phi_*$.
In these cases, with $\dot \phi_i = 0$, one again obtains a very restricted family of behaviors
for $w(a)$:  the functional form for $w(a)$ contains one free parameter, determined
by the value of $V''(\phi_*)$, but otherwise depends only on $\Omega_{\phi 0}$ and $w_0$.
These results were further generalized by Chiba \cite{Chiba}, who expanded $V(\phi)$ up
to second order around the initial value for $\phi$, rather than assuming that
$\phi$ begins near an extremum.  (See also Ref. \cite{Huiyiing} for evolution
near an inflection point, for which $V(\phi)$ is approximately cubic).

This paper represents the logical extension of Refs. \cite{ds1,ds3}.  As in Refs. \cite{ds1,ds3}, we assume that the potential
can be well-approximated by Eq. (\ref{quadratic}), but we introduce one additional degree of freedom:  the
initial value of $\dot \phi$ is allowed to be nonzero.  This results in a much wider range of behaviors
for $w(a)$ than in Refs. \cite{ds1,ds3}, but, as we will see, the functional form for $w(a)$ is not arbitrary.  Instead, there is a
well-defined set of functions $w(a)$ that depends on $\Omega_{\phi0}$, $w_0$, $V''(\phi_*)$, and the new
parameter we introduce here, $\dot \phi_i$, or, equivalently, $w_i$.

The most important model that is well-described by Eq. (\ref{quadratic})
is the PNGB model \cite{Frieman,Dutta,Albrecht,LinderPNGB}, for which
\begin{equation}
V(\phi) = M^4[\cos(\phi/f)+1].
\end{equation}
However, {\it any} potential with a local maximum or minimum can
be Taylor-expanded in the form of Eq. (\ref{quadratic}), and our purpose
is to show that all such models converge to a common set of behaviors for
$w(a)$.

In the next section, we derive our approximation for $w(a)$ given the assumptions outlined here.  In Sec.
III, we compare our results to numerical integration of the scalar field equation of motion, and find
excellent agreement.  In Sec. IV, we discuss a general limit on the turning points
of $w(a)$ that is particularly useful for some of the potentials considered in this paper.  Limits
from observational data are presented in Sec. V.
Our conclusions are summarized in Sec. VI.

\section{The Derivation}

\subsection{$K^2 > 0$}

Since we are interested in the evolution of the universe at late times, we assume a
flat universe containing only matter and a scalar
field.  Also, we restrict our attention to
quintessence models in which $w \approx -1$ so
that we can make use of the relation:
\begin{equation}
\rho_\phi \approx \rho_{\phi 0} \approx -p_\phi,
\end{equation}
where the density and pressure of the scalar field are given by, respectively,
\begin{equation}
\label{density}
\rho_\phi={\frac{1}{2}}{\dot{\phi}}^2+V(\phi),
\end{equation}
and
\begin{equation}
p_\phi={\frac{1}{2}}{\dot{\phi}}^2 - V(\phi).
\end{equation}
We take $\hbar = c = 8\pi G = 1$ throughout.
The equation of motion for the scalar field is
\begin{equation}
\label{motion}
\ddot{\phi}+3H\dot{\phi}+\frac{dV}{d\phi}=0,
\end{equation}
where the Hubble parameter $H$ is given by
\begin{equation}
\label{H2}
H^2 = \frac{1}{3}\rho_T,
\end{equation}
and the subscript $T$ refers to the total (matter plus scalar field) density.

Our derivation follows Ref. \cite {ds1}. We make the change of variables
\begin{equation}
\label{udef}
u=(\phi-\phi_*) a^{3/2},
\end{equation}
where $a$ is the scale factor and
$\phi_*$ is a local maximum or minimum of the potential.
Then Eq. (\ref{motion}) becomes
\begin{equation}
\label{ueq}
\ddot{u} + \frac{3}{4}p_Tu + a^{3/2}\frac{dV}{d\phi} = 0.
\end{equation}
Since the matter component is pressureless, we can make the approximation
\begin{equation}
p_T = p_\phi \approx - \rho_{\phi 0}.
\end{equation}
Furthermore, we assume that $V(\phi)$ can be expanded about
the extremum $\phi_*$ as
\begin{equation}
\label{expand}
V(\phi) \approx V(\phi_*) + \frac{1}{2}V''(\phi_*),
(\phi-\phi_*)^2
\end{equation}
and $\rho_{\phi_0} \approx V(\phi_*)$.
Then Eq. (\ref{ueq}) becomes:
\begin{equation}
\label{ufinal}
\ddot{u}+[V''(\phi_*) - \frac{3}{4}V(\phi_*)]u = 0,
\end{equation}
which can be easily solved for $u(t)$.

Eq. (\ref{ufinal}) has the general solution:
\begin{equation}
u=Ce^{kt}+De^{-kt},
\end{equation}
where
\begin{equation}
k=\sqrt{\frac{3}{4}V(\phi_*) - V''(\phi_*)}.
\end{equation}
Then, taking the derivative and dividing by $k$, we have:
\begin{equation}
\frac{\dot{u}}{k}=Ce^{kt}-De^{-kt}.
\end{equation}
From here, we solve for $C$ and $D$ as follows:
\begin{equation}
C=\frac{ku+\dot{u}}{2k}e^{-kt},
\end{equation}
\begin{equation}
D=\frac{ku-\dot{u}}{2k}e^{kt}.
\end{equation}
Taking $u$ from Eq. (\ref{udef}) and
\begin{equation}
\dot{u}=\dot{\phi}a^{3/2}+\frac{3}{2}(\phi-\phi_*)Ha^{3/2},
\end{equation}
we can express $C$ and $D$ in terms of the initial conditions $\phi_i$
and $\dot{\phi_i}$.  To further simplify our expressions, it is convenient
to take the extremum of $V$ to be at $\phi_* = 0$.  Then we obtain:
\begin{equation}
C=\frac{k\phi_i+\dot{\phi_i}+\frac{3}{2}\phi_i H_i}{2k}a_i^{3/2}e^{-kt_i},
\end{equation}
and
\begin{equation}
D=\frac{k\phi_i-\dot{\phi_i}-\frac{3}{2}\phi_i H_i}{2k}a_i^{3/2}e^{kt_i},
\end{equation}
and our general solution for $\phi$ becomes
\begin{widetext}
\begin{equation}
\label{phi(a,t)}
\phi=\frac{1}{2k}\left(\frac{a}{a_i}\right)^{-3/2}
\left\{\left[\dot{\phi_i}+\left(\frac{3}{2}H_i+k\right)\phi_i\right]e^{k(t-t_i)}\right.
-\left.\left[\dot{\phi_i}+\left(\frac{3}{2}H_i-k\right)\phi_i\right]e^{-k(t-t_i)}
\right\}.
\end{equation}
\end{widetext}

To make further progress, we must make some assumption for the functional
form of $a(t)$.  If $w \approx -1$, then we expect
$a(t)$ to be well-approximated by the expression
corresponding to a $\Lambda$CDM universe, namely
\begin{equation}
\label{a(t)}
a(t)=\left(\Omega_{\phi 0}^{-1}-1\right)^{1/3}\sinh^{2/3}(t/t_\Lambda),
\end{equation}
where $t_\Lambda$ is defined to be:
\begin{equation}
t_\Lambda = 2/\sqrt{3\rho_{\phi 0}},
\end{equation}
and $a=1$ corresponds to the present.
We expect Eq. (\ref{a(t)}) to be an excellent approximation
as long as $w$ remains close to $-1$.  Then the Hubble parameter is given by
\begin{equation}
H(t)=\frac{2}{3t_\Lambda}\coth(t/t_\Lambda),
\end{equation} 
and Eq. (\ref{phi(a,t)}) becomes
\begin{widetext}
\begin{equation}
\label{phi1(t)}
\phi(t)=\frac{1}{2k}\frac{\sinh(t_i/t_\Lambda)}{\sinh(t/t_\Lambda)}
\left\{\left[\dot{\phi_i}+\left(\frac{\coth(t_i/t_\Lambda)}{t_\Lambda}
+k\right)\phi_i\right]e^{k(t-t_i)}\right. \\ 
-\left.\left[\dot{\phi_i}+\left(
\frac{\coth(t_i/t_\Lambda)}{t_\Lambda}-k\right)\phi_i\right]e^{-k(t-t_i)}
\right\}
\end{equation}
\end{widetext}
If we take $t_i \rightarrow 0$ and $\dot \phi_i = 0$, we regain the corresponding
expression for $\phi(t)$ from Ref. \cite{ds1}: 
\begin{equation}
\label{phi(t)}
\phi(t) = \frac{\phi_i}{kt_\Lambda}\frac{\sinh(kt)}{\sinh(t/t_\Lambda)}.
\end{equation}

The physically measurable quantity of interest is not $\phi$, but the equation of state parameter $w$,
given by
\begin{equation}
1+w = \frac{\dot{\phi}^2}{\rho_{\phi}}.
\end{equation}
In the limit where $w \approx -1$, this is well-approximated by
\begin{eqnarray}
1+w&=&\frac{\dot{\phi}^2}{\rho_{\phi 0}},\\
\label{wapprox}
&=&\frac{3}{4}\dot{\phi}^2 t_\Lambda^2.
\end{eqnarray}
Using Eq. (\ref{phi(t)}) to derive $\dot{\phi}$, we obtain:
\begin{multline}
1+w(t)=\frac{3}{4}\left\{\left[\frac{\phi_i}{kt_\Lambda}\cosh(t_i/t_\Lambda)+\frac{\dot{\phi_i}}{k}
\sinh(t_i/t_\Lambda)\right]\right. \\
\times \left[\frac{kt_\Lambda\cosh(k(t-t_i))\sinh(t/t_\Lambda)
-\sinh(k(t-t_i))\cosh(t/t_\Lambda)}{\sinh^2(t/t_\Lambda)}\right] \\
+\left.\phi_i\sinh(t_i/t_\Lambda)\left[\frac{kt_\Lambda\sinh(k(t-t_i))
\sinh(t/t_\Lambda)-\cosh(k(t-t_i))\cosh(t/t_\Lambda)}{\sinh^2(t/t_\Lambda)}
\right]\right\}^2
\end{multline}
Measurements of $w$ are derived as a function of redshift, or equivalently, as a function of $a$.
Inverting Eq. (\ref{a(t)}) to derive $t(a)$, our expression for $1+w$ becomes
\begin{multline}
\label{1+wbegin}
1+w(a)=\frac{3}{16K^2}\left(\frac{a}{a_i}\right)^{3(K-1)}\left\{\left(K-F(a)\right)\left[\dot{\phi_i}t_\Lambda+\phi_i(F(a_i)+K)\right]\left(\frac{F(a)+1}{F(a_i)+1}\right)^K\right. \\
+\left.\left(K+F(a)\right)\left[\dot{\phi_i}t_\Lambda+\phi_i(F(a_i)-K)\right]\left(\frac{F(a)-1}{F(a_i)-1}\right)^K\right\}^2,
\end{multline}
where the function $F(a)$ is defined as
\begin{equation}
\label{F(a)}
F(a)=\sqrt{1+(\Omega_{\phi 0}^{-1}-1)a^{-3}},
\end{equation}
and 
\begin{equation}
K = k t_\Lambda,
\end{equation}
which can be written in terms of the quintessence potential as
\begin{equation}
\label{Kdef}
K = \sqrt{1-(4/3) V^{\prime \prime}(\phi_*)/V(\phi_*)}.
\end{equation}

We would like to write Eq. (\ref{1+wbegin}) in terms of $w_i$ and $w_0$ instead of $\phi_i$ and $\dot \phi_i$.  We first make
the substitution
\begin{equation}
\label{phitow}
\dot{\phi_i}t_\Lambda = \pm\sqrt{\frac{4}{3}(1+w_i)},
\end{equation}
which gets rid of $\dot \phi_i$.  To eliminate $\phi_i$, we take $a = a_0 =1$
in Eq. (\ref{1+wbegin}), solve for $\phi_i$, and then substitute this expression for $\phi_i$ back into Eq. (\ref{1+wbegin}).  This gives
an expression for $1+w(a)$ in terms of $w_i$, $w_0$, $\Omega_{\phi 0}$, and $V^{\prime \prime}(\phi_*)$ (expressed in terms of $K$):
\begin{equation}
\label{1+w}
1+w(a)=\left(\frac{X_K(a)}{X_K(a_0)}\sqrt{1+w_0}\pm\frac{Y_K(a)}{Y_K(a_i)}\sqrt{1+w_i}\right)^2,
\end{equation}
where
\begin{multline}
X_K(a)=\left(\frac{a}{a_i}\right)^{\frac{3}{2}K}\left(\frac{a}{a_0}\right)^{-\frac{3}{2}}\left[\left(F(a_i)+K\right)\left(K-F(a)\right)\left(\frac{F(a)+1}{F(a_i)+1}\right)^K \right. \\
+\left.\left(F(a_i)-K\right)\left(K+F(a)\right)\left(\frac{F(a)-1}{F(a_i)-1}\right)^K\right],
\end{multline}
and
\begin{multline}
\label{YaK}
Y_K(a)=\left(\frac{a}{a_0}\right)^{\frac{3}{2}K}\left(\frac{a}{a_i}\right)^{-\frac{3}{2}}\left[\left(F(a_0)+K\right)\left(K-F(a)\right)\left(\frac{F(a)+1}{F(a_0)+1}\right)^K \right. \\
+\left.\left(F(a_0)-K\right)\left(K+F(a)\right)\left(\frac{F(a)-1}{F(a_0)-1}\right)^K\right].
\end{multline}
The two solutions in Eq. (\ref{1+w}) arise because $\dot \phi_i$ is not uniquely determined by $w_i$ in
Eq. (\ref{phitow}); they correspond to a scalar field initially rolling either uphill or downhill in the potential.

An interesting limiting case is $a_i \rightarrow 0$, which corresponds to the case considered
in Ref. \cite{ds1}.
In this limit, $F(a_i) \rightarrow \infty$, and we obtain
\begin{equation}
\frac{X_K(a)}{X_K(a_0)}=a^{\frac{3}{2}(K-1)}\left[\frac{\left(K-F(a)\right)
\left(F(a)+1\right)^K+\left(K+F(a)\right)\left(F(a)-1\right)^K}{\left(K-\Omega_{\phi  0}^{-1/2}\right)\left(\Omega_{\phi  0}^{-1/2}+1\right)^K
+\left(K+\Omega_{\phi  0}^{-1/2}\right)\left(\Omega_{\phi  0}^{-1/2}-1\right),
^K}\right]
\end{equation}
and
\begin{equation}
\frac{Y_K(a)}{Y_K(a_i)} = 0.
\end{equation}
Thus, we see that as $a_i \rightarrow 0$, Eq. (\ref{1+w}) becomes the previously found
Dutta-Scherrer \cite{ds1} solution for hilltop quintessence, as expected.

The solution given by Eqs. (\ref{1+w})$-$(\ref{YaK}) applies only to the case where $K^2 > 0$ (i.e., the same case considered in Ref. \cite{ds1}).
Now we extend this solution to all values of $K$.

\subsection{$K^2 < 0$}

Consider first the case where $K^2 < 0$.
Technically, if we write $K$ as $i\kappa t_\Lambda$, where $\kappa$ is a positive real, our previous
results give the correct formula for $1+w(a)$.  However, the resulting expressions are rather opaque.
Instead, we solve Eq. (\ref{ufinal}) for $K^2<0$, giving
\begin{equation}
u=A\sin(\kappa t)+B\cos(\kappa t),
\end{equation}
where
\begin{equation}
\kappa = \sqrt{V^{\prime\prime}(\phi_*) - \frac{3}{4} V(\phi_*)}.
\end{equation}
As before, we can solve for $A$ and $B$ in terms of $\phi_i$ and
$\dot \phi_i$:
\begin{equation}
A =\left(\phi_i\sin(\kappa t_i)+\frac{\dot{\phi_i}+\frac{3}{2}\phi_i H_i}{\kappa}\cos(\kappa t_i)\right)a_i^{3/2},
\end{equation}
and
\begin{equation}
B =\left(\phi_i\cos(\kappa t_i)-\frac{\dot{\phi_i}+\frac{3}{2}\phi_i H_i}{\kappa}\sin(\kappa t_i)\right)a_i^{3/2},
\end{equation}
and the equation for $\phi(t)$ (analogous to Eq. \ref{phi1(t)}) is
\begin{multline}
\label{phiK<0}
\phi(t)=\frac{1}{\kappa}\left(\frac{\sinh(t_i/t_\Lambda)}{\sinh(t/t_\Lambda)}\right)
\left\{\left[\phi_i\kappa\sin(\kappa t_i)+\left(\dot{\phi_i}+\frac{3}{2}\phi_i H_i\right)\cos(\kappa t_i)\right]\sin(\kappa t)\right. \\
+\left.\left[\phi_i\kappa\cos(\kappa t_i)-\left(\dot{\phi_i}+\frac{3}{2}\phi_i H_i\right)\sin(\kappa t_i)\right]\cos(\kappa t)\right\}
\end{multline}

As before, we use Eq. (\ref{wapprox}) to derive $w$ as a function of $\dot \phi$.  Taking the derivative of Eq. (\ref{phiK<0})
and following the procedure of Sec. IIA to rewrite $\phi_i$ and $\dot \phi_i$ in terms of $w_0$ and $w_i$, we obtain the $K^2<0$ expression
for $1+w$:
\begin{equation}
\sqrt{1+w(a)}=\frac{X_K(a)}{X_K(a_0)}\sqrt{1+w_0}\pm\frac{Y_K(a)}{Y_K(a_i)}\sqrt{1+w_i},
\end{equation}
where
\begin{multline}
X_K(a)=\left(\frac{a}{a_0}\right)^{-3/2}
\left\{
\left[\left|K\right|\sin\left(\left|K\right|\ln\sqrt{\frac{F(a_i)+1}{F(a_i)-1}}\right)+F(a_i)\cos\left(\left|K\right|\ln\sqrt{\frac{F(a_i)+1}{F(a_i)-1}}\right)\right]\right. \\
\times\left[\left|K\right|\cos\left(\left|K\right|\ln\sqrt{\frac{F(a)+1}{F(a)-1}}\right)-F(a)\sin\left(\left|K\right|\ln\sqrt{\frac{F(a)+1}{F(a)-1}}\right)\right] \\
-\left[\left|K\right|\cos\left(\left|K\right|\ln\sqrt{\frac{F(a_i)+1}{F(a_i)-1}}\right)-F(a_i)\sin\left(\left|K\right|\ln\sqrt{\frac{F(a_i)+1}{F(a_i)-1}}\right)\right] \\
\times\left.\left[\left|K\right|\sin\left(\left|K\right|\ln\sqrt{\frac{F(a)+1}{F(a)-1}}\right)+F(a)\cos\left(\left|K\right|\ln\sqrt{\frac{F(a)+1}{F(a)-1}}\right)\right]
\right\},
\end{multline}
and
\begin{multline}
Y_K(a)=\left(\frac{a}{a_i}\right)^{-3/2}
\left\{
\left[\left|K\right|\cos\left(\left|K\right|\ln\sqrt{\frac{F(a_0)+1}{F(a_0)-1}}\right)-F(a_0)\sin\left(\left|K\right|\ln\sqrt{\frac{F(a_0)+1}{F(a_0)-1}}\right)\right]\right. \\
\times\left[\left|K\right|\sin\left(\left|K\right|\ln\sqrt{\frac{F(a)+1}{F(a)-1}}\right)+F(a)\cos\left(\left|K\right|\ln\sqrt{\frac{F(a)+1}{F(a)-1}}\right)\right] \\
-\left[\left|K\right|\sin\left(\left|K\right|\ln\sqrt{\frac{F(a_0)+1}{F(a_0)-1}}\right)+F(a_0)\cos\left(\left|K\right|\ln\sqrt{\frac{F(a_0)+1}{F(a_0)-1}}\right)\right] \\
\times\left.\left[\left|K\right|\cos\left(\left|K\right|\ln\sqrt{\frac{F(a)+1}{F(a)-1}}\right)-F(a)\sin\left(\left|K\right|\ln\sqrt{\frac{F(a)+1}{F(a)-1}}\right)\right]
\right\}.
\end{multline}

\subsection{$K^2=0$}

Now we take $K^2 = 0$.
We first solve Eq. (\ref{ufinal}) in the $k=0$ case to find that
\begin{equation}
u =At+B.
\end{equation}
Evaluating at $t_i$, we solve for $A$ and $B$ to obtain
\begin{equation}
A=\dot{u}=\left(\dot{\phi_i}+\frac{3}{2}H_i \phi_i\right)a_i^{3/2},
\end{equation}
and
\begin{equation}
B=\left[\phi_i+\left(\dot{\phi_i}+\frac{3}{2}H_i \phi_i\right)t_i\right]a_i^{3/2},
\end{equation}
so
\begin{equation}
\phi=\left[\phi_i+\left(\dot{\phi_i}+\frac{3}{2}H_i \phi_i\right)\left(t-t_i\right)\right]\left(\frac{a}{a_i}\right)^{-3/2}.
\end{equation}
Then using the $\Lambda$CDM approximation for $a(t)$ (Eq. \ref{a(t)}), we have
\begin{equation}
\phi(t)=\left[\phi_i+\left(\dot{\phi_i}+\frac{3}{2}H_i \phi_i\right)\left(t-t_i\right)\right]\left(\frac{\sinh(t_i/t_\Lambda)}{\sinh(t/t_\Lambda)}\right).
\end{equation}
We use the same procedure as in the previous two cases to express $1+w$ as a function of $\dot \phi$, and eliminate $\phi_i$
and $\dot \phi_i$ in favor of $w_0$ and $w_i$, yielding the result:
\begin{equation}
\sqrt{1+w(a)}=\frac{X_0(a)}{X_0(a_0)}\sqrt{1+w_0}\pm\frac{Y_0(a)}{Y_0(a_i)}\sqrt{1+w_i},
\end{equation}
where
\begin{equation}
X_0=\left(\frac{a}{a_0}\right)^{-3/2}\left[F(a_i)-F(a)-\frac{1}{2}F(a_i)F(a)\ln\left[\left(\frac{F(a_i)-1}{F(a_i)+1}\right)\left(\frac{F(a)+1}{F(a)-1}\right)\right]\right],
\end{equation}
and
\begin{equation}
Y_0=\left(\frac{a}{a_i}\right)^{-3/2}\left[F(a_0)-F(a)-\frac{1}{2}F(a_0)F(a)\ln\left[\left(\frac{F(a_0)-1}{F(a_0)+1}\right)\left(\frac{F(a)+1}{F(a)-1}\right)\right]\right].
\end{equation}

\subsection{Combined Solution}

Because of the high degree of symmetry in our expressions for $1+w$, we can write them in a simpler way that combines all three solutions,
namely
\begin{multline}
\label{wfinal}
\sqrt{1+w(a)}=\left(\frac{a}{a_0}\right)^{-3/2}\left(\frac{f_K(a)g_K(a_i)-f_K(a_i)g_K(a)}{f_K(a_0)g_K(a_i)-f_K(a_i)g_K(a_0)}\right)\sqrt{1+w_0} \\
\pm\left(\frac{a}{a_i}\right)^{-3/2}\left(\frac{f_K(a)g_K(a_0)-f_K(a_0)g_K(a)}{f_K(a_i)g_K(a_0)-f_K(a_0)g_K(a_i)}\right)\sqrt{1+w_i}
\end{multline}
where $f_K(a)$ and $g_K(a)$ are the much more manageable functions:
\begin{equation}
\label{fK}
f_K(a)=\left\{ 
\begin{array}{ll}
\left(K+F(a)\right)\left(\frac{F(a)-1}{F(a)+1}\right)^{K/2} & (K^2>0) \\
F(a) & (K^2=0) \\
|K|\sin\ln\left(\frac{F(a)+1}{F(a)-1}\right)^{|K|/2} + F(a)\cos\ln\left(\frac{F(a)+1}{F(a)-1}\right)^{|K|/2} & (K^2<0)
\end{array}
\right.
\end{equation}
\begin{equation}
\label{gK}
g_K(a)=\left\{ 
\begin{array}{ll}
\left(K-F(a)\right)\left(\frac{F(a)+1}{F(a)-1}\right)^{K/2} & (K^2>0) \\
1-F(a)\ln\left(\frac{F(a)+1}{F(a)-1}\right)^{1/2} & (K^2=0) \\
|K|\cos\ln\left(\frac{F(a)+1}{F(a)-1}\right)^{|K|/2} - F(a)\sin\ln\left(\frac{F(a)+1}{F(a)-1}\right)^{|K|/2} & (K^2<0)
\end{array}
\right.
\end{equation}
Eqs. (\ref{wfinal})-(\ref{gK}) are our main result, with $F(a)$ given by Eq. (\ref{F(a)}), and $K$ given by Eq. (\ref{Kdef}).
The two signs in Eq. (\ref{wfinal}) correspond to the two different possible initial directions of motion
for $\phi$ for a given value of $w_i$.

\section{Comparison with Exact Solutions}

In the following section, we compare our approximation for $w(a)$ with numerical results for
three different potentials: a quadratic
\begin{equation}
\label{Vquadratic}
V(\phi)=V_0+V_2\phi^2,
\end{equation}
a Gaussian
\begin{equation}
\label{VGaussian}
V(\phi)=V_0e^{-\phi^2/\sigma^2},
\end{equation}
and the aforementioned PNGB potential
\begin{equation}
\label{VPNGB}
V(\phi)=V_0\left[1+\cos\left(\phi/f\right)\right].
\end{equation}
These, of course, are not meant to be an exhaustive list of scalar field potentials,
but we merely wish to show that these different potentials will produce similar evolution
for $w(a)$, and that this evolution agrees with our analytic approximation.
Our numerical solutions to Eqs. (\ref{motion}) and (\ref{H2}) are constrained
by four boundary conditions: $w_i$, $w_0$, $\Omega_{\phi 0}$, and $K$, which
together determine the parameters in each of our potentials. The definition of $K$ requires that
\begin{equation}
1-K^2=\frac{4V''(0)}{3V(0)},
\end{equation}
and so
\begin{equation}
V''(0)=\frac{3}{4}(1-K^2)V(0).
\end{equation}
Furthermore, through the condition on $\Omega_{\phi 0}$, we have:
\begin{equation}
\rho_{\phi 0}=\frac{1}{2}\dot \phi_0^2+V(\phi_0),
\end{equation}
and since $\dot \phi_0^2=(1+w_0)\rho_{\phi 0}$ and $\rho_{\phi 0}=3H_0^2\Omega_{\phi 0}$, this expression can be rewritten as:
\begin{equation}
V(\phi_0)=\frac{3}{2}H_0^2\Omega_{\phi_0}\left(1-w_0\right).
\end{equation}

These conditions can then be applied to the various potentials to determine the parameters for each case.  For the
quadratic, they become
\begin{equation}
\label{quadK}
V_2=\frac{3}{8}(1-K^2)V_0,
\end{equation}
and
\begin{equation}
V_0 + V_2 \phi_0^2=\frac{3}{2}H_0^2\Omega_{\phi 0}(1-w_0),
\end{equation}
respectively, which yields
\begin{equation}
V_0=\frac{\frac{3}{2}H_0^2\Omega_{\phi 0}(1-w_0)}{1+\frac{3}{8}(1-K^2)\phi_0^2},
\end{equation}
\begin{equation}
V_2=\frac{3}{8}(1-K^2)\frac{\frac{3}{2}H_0^2\Omega_{\phi 0}(1-w_0)}{1+\frac{3}{8}(1-K^2)\phi_0^2}.
\end{equation}
Similarly, we have for the Gaussian
\begin{equation}
\sigma^2=\frac{8}{3(K^2-1)}
\end{equation}
\begin{equation}
V_0=\frac{3}{2}H_0^2\Omega_{\phi_0}(1-w_0)e^{\frac{3}{8}(K^2-1)\phi_0^2}.
\end{equation}
This corresponds to a Gaussian function only for $K^2 > 1$.  For $K^2 < 1$, we have
$\sigma^2 < 0$.  While this is no longer a Gaussian,
the resulting potential has a local minimum at $\phi =0$ and is nonetheless valid for our purposes.  In fact, this potential
represents a special case of the SUGRA-inspired potentials proposed in Ref. \cite{Brax}.

For a PNGB potential
\begin{equation}
f=\frac{2}{3\sqrt{K^2-1}},
\end{equation}
\begin{equation}
V_0=\frac{\frac{3}{2}H_0^2\Omega_{\phi 0}(1-w_0)}{1+\cos(\frac{3}{2}\phi_0\sqrt{K^2-1})}.
\end{equation}
This gives a PNGB form of the potential only for $K^2 > 1$, for evolution near
the maximum of the potential.  There is no obvious analog in the case of evolution
near the potential minimum ($K^2 < 1$), so we simply use the imaginary value of $\sqrt{K^2-1}$ in
this case, producing a hyperbolic cosine potential well.

Thus, the appropriate parameters for a potential with a given value of $K$ can
be determined by specifying $w_0$ and $\phi_0$.  Three of the four boundary
conditions, namely $K$, $\Omega_{\phi 0}$, and $w_0$, therefore depend only on
the present state, and hence it is appropriate to run the simulation backwards
from $a_0$ to $a_i$ in order to find a value of $\phi_0$ which satisfies the
fourth boundary condition: $w_i$.  The particular algorithm in these simulations
first finds values of $\phi_0$ between which $\dot{\phi_i}$ changes sign, and
then pinpoints the $\phi_0$ value for which $w_i=-1$.  Then, by deviating from
this point, appropriate $\phi_0$ values can be found corresponding to the correct $w_i$.

\begin{figure}[!t]
\centerline{\epsfxsize=4truein\epsffile{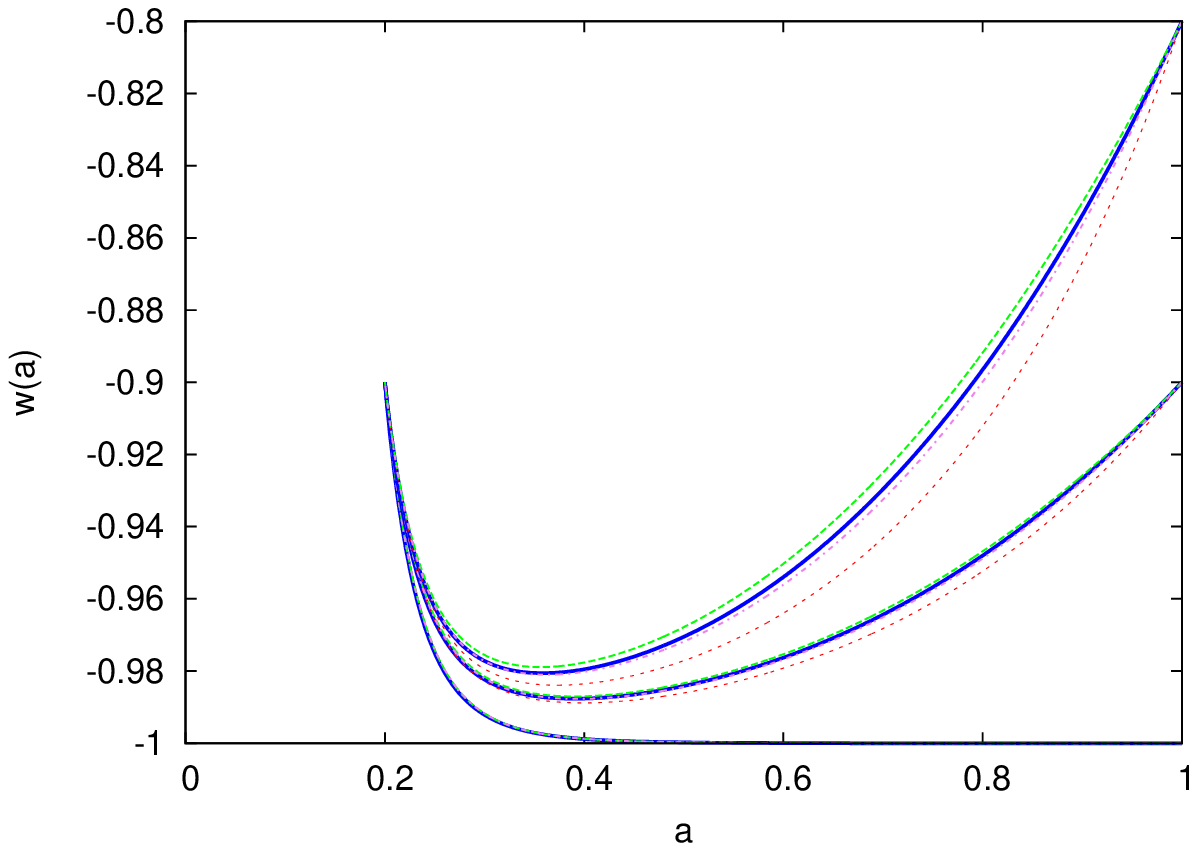}}
\caption{Evolution of $w(a)$ for $K^2=4$, fixed $w_i$ and $a_i$, and several different values of $w_0$.  Solid
(blue) curve gives our ($+$) analytic approximation (Eqs. \ref{wfinal}$-$\ref{gK}).  Dashed curves give the exact
(numerical) evolution for (top to bottom) the Gaussian (green), PNGB (violet), and quadratic (red) potentials.}
\end{figure}

\begin{figure}[!t]
\centerline{\epsfxsize=4truein\epsffile{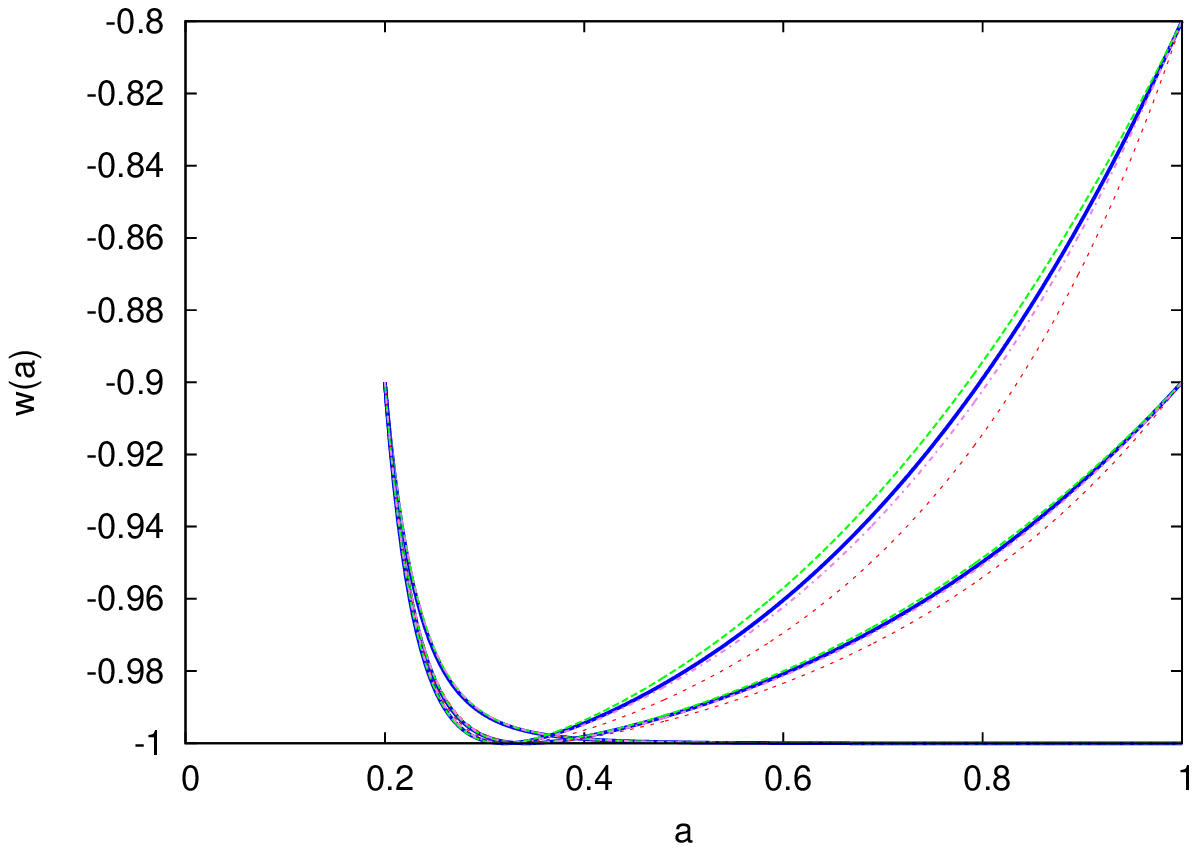}}
\caption{As Fig. 1, with the ($-$) solution for $K^2 = 4$.}
\end{figure}

\begin{figure}[!t]
\centerline{\epsfxsize=4truein\epsffile{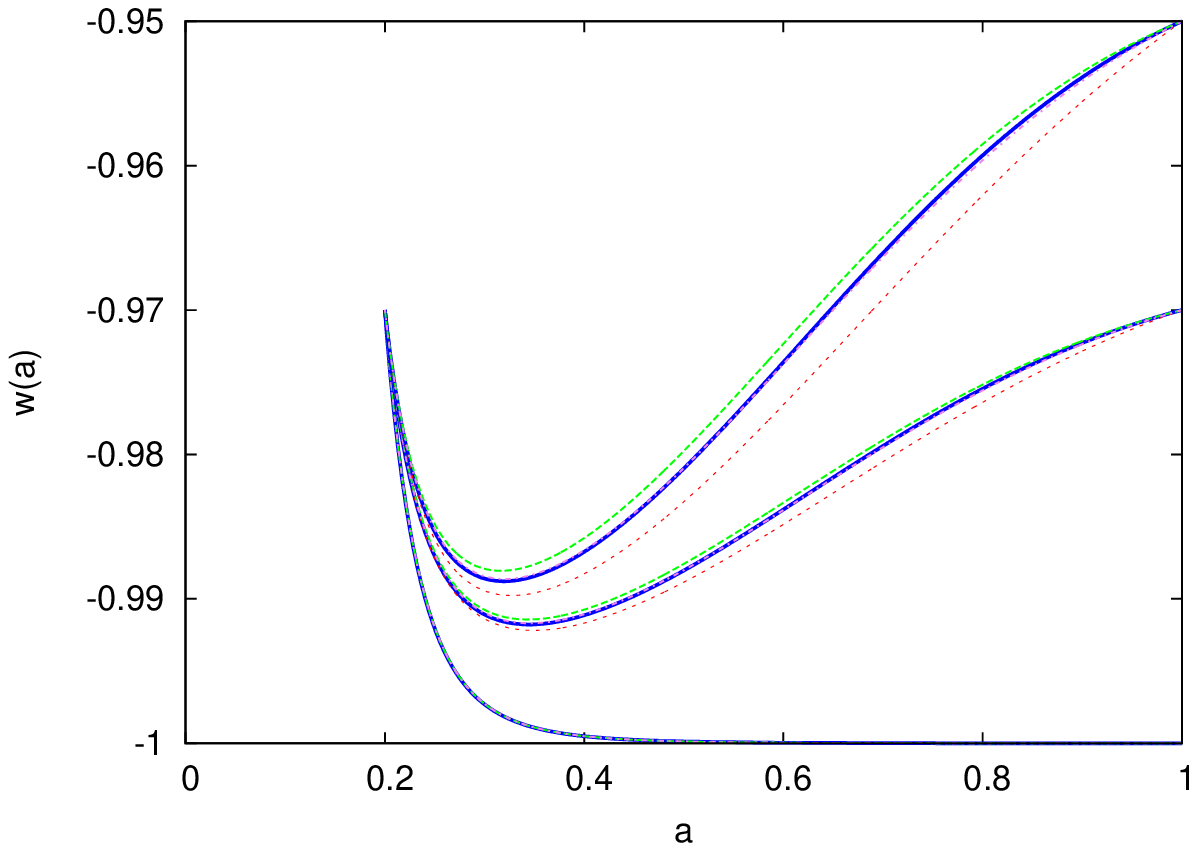}}
\caption{As Fig. 1, with the ($+$) solution for $K^2=0$.}
\end{figure}

\begin{figure}[!t]
\centerline{\epsfxsize=4truein\epsffile{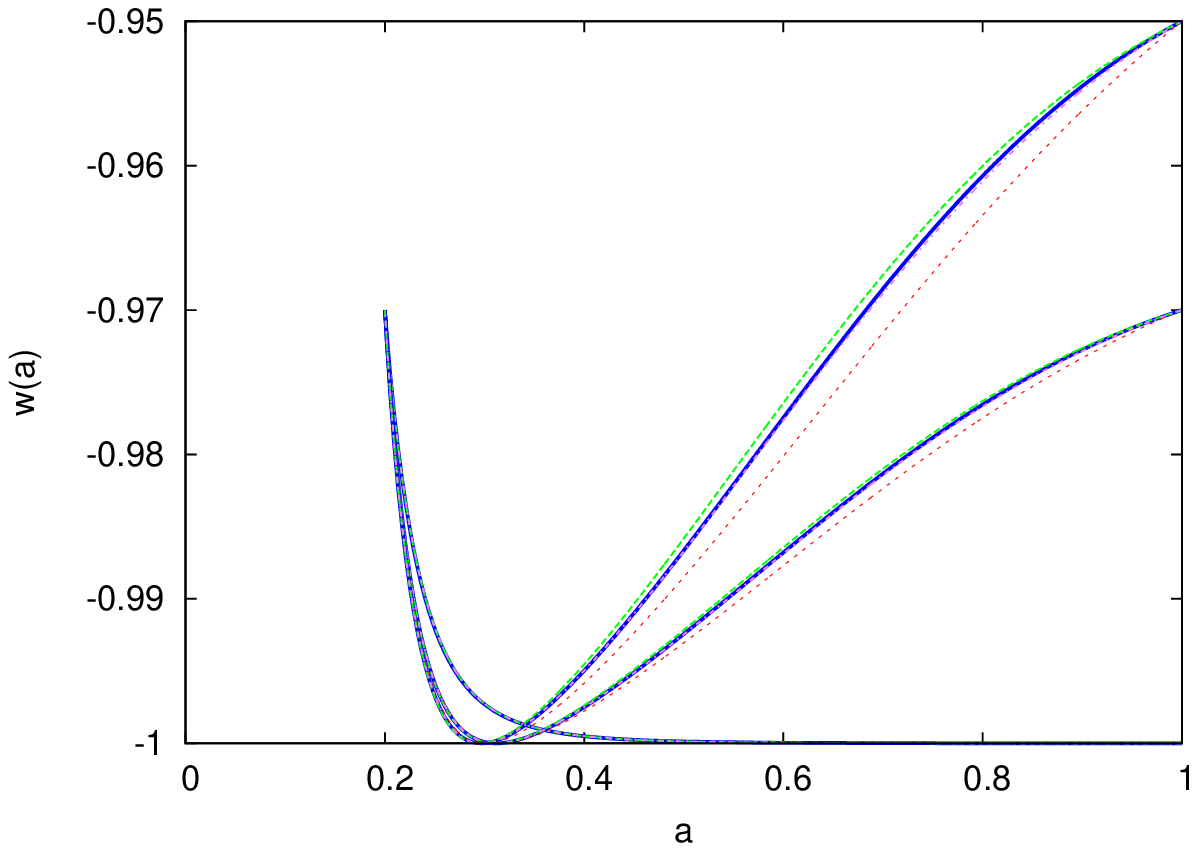}}
\caption{As Fig. 1, with the ($-$) solution for $K^2 = 0$.}
\end{figure}

\begin{figure}[!t]
\centerline{\epsfxsize=4truein\epsffile{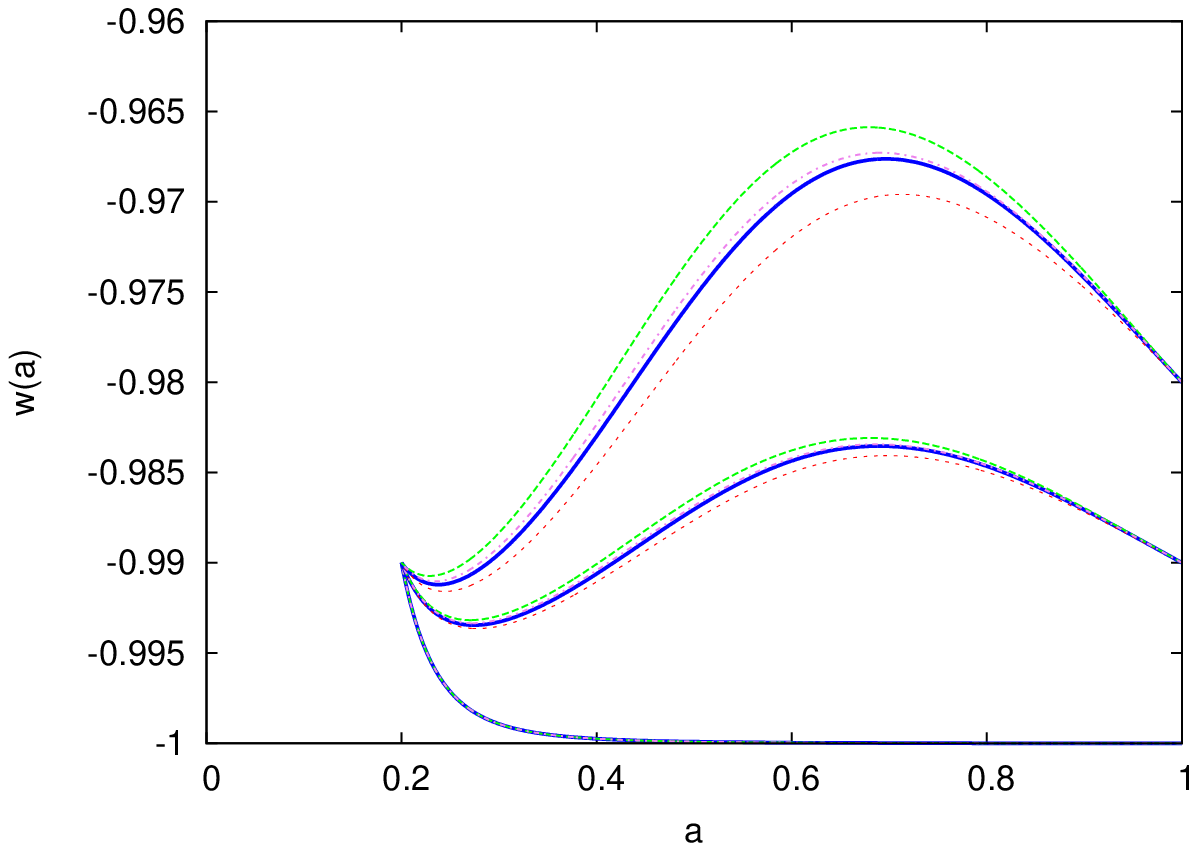}}
\caption{As Fig. 1, with the ($+$) solution for $K^2=-4$.}
\end{figure}

\begin{figure}[!t]
\centerline{\epsfxsize=4truein\epsffile{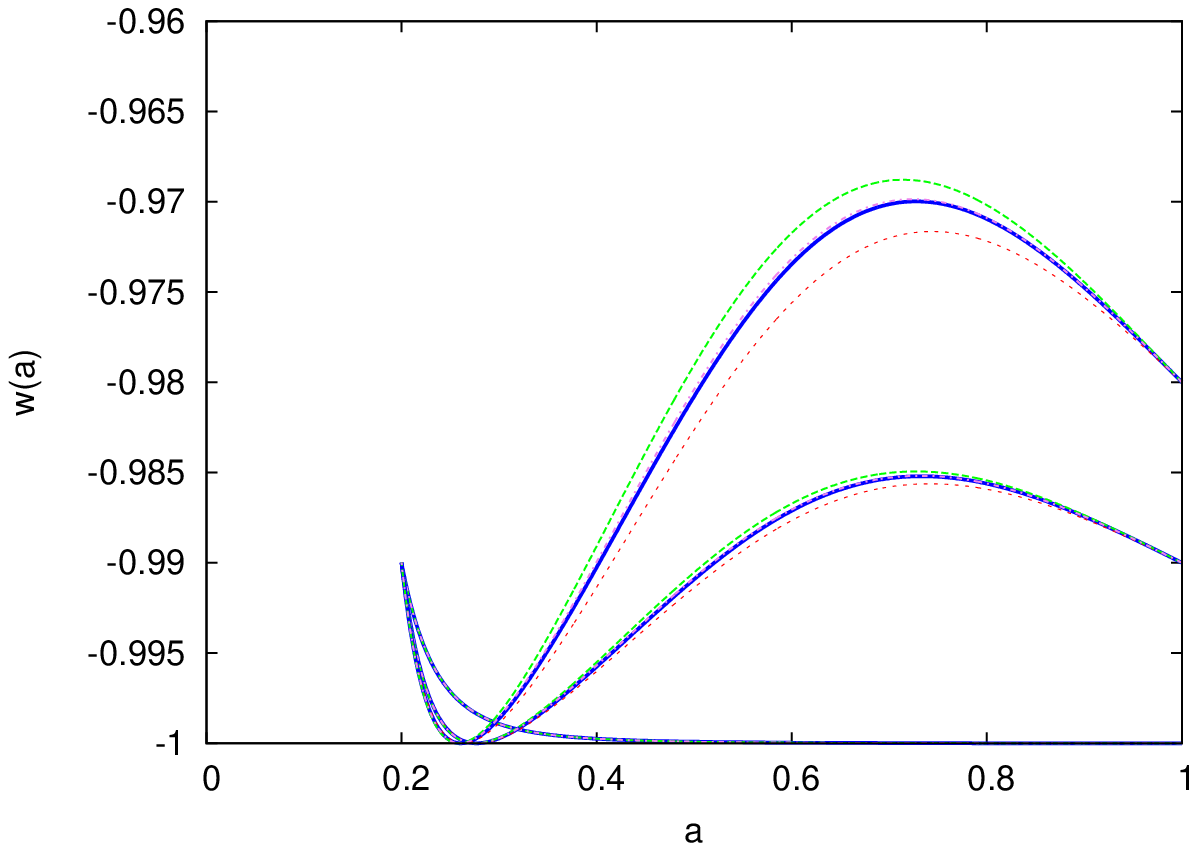}}
\caption{As Fig. 1, with the ($-$) solution for $K^2 = -4$.}
\end{figure}

Using this method, numerical solutions were obtained for three values of $K^2$:  $-4$, $0$, and $+4$, and for both initial directions
of motion of the field, corresponding to the $+$ and $-$ solutions in Eq. (\ref{wfinal}).
The value of $a_i$ was chosen, somewhat arbitrarily, to be $0.2$, corresponding to $z=4$,
to highlight the evolution of $w$ between $a_i$ and the present, which is taken
to be at $a=1$ with $\Omega_{\phi 0} = 0.73$.

Our results are illustrated in Figs. 1-6.  They all show excellent agreement between the exact numerical evolution and our analytic
expressions. 
The general behavior of these functions is similar for all six cases:  the field begins by freezing, with $w$ decreasing to a value
close to $-1$, followed by thawing (i.e., increasing $w$) as the field rolls downhill in the potential.  The $K^2 = 4$ cases correspond
to oscillatory evolution of $\phi$, so that $w$ eventually reaches a maximum value as it rolls through the minimum in the potential and then
decreases when it rolls uphill on the other side.  Even larger values of $K^2$ are found to result in multiple oscillations.

\section{Another constraint on the functional form for $w(a)$}

Here we briefly discuss a general constraint on the evolution of $w$ that
is particularly useful for some of the models in this paper.  The equation for $dw/da$ can be written in the form
\cite{SteinhardtWangZlatev,Linder}
\begin{equation}
a \frac{dw}{da} = -3(1-w)(1+w) - \frac{V^\prime}{V}(1-w)\sqrt{3(1+w)\Omega_\phi}.
\end{equation}
This expression allows us to determine the location of turning points in $w(a)$, at which
$w$ takes on a maximum or minimum value.  Setting the right-hand side to zero, we obtain
\begin{equation}
\label{wvalue}
1+w_{m} = \frac{1}{3}\left(\frac{V^\prime}{V}\right)^2 \Omega_\phi,
\end{equation}
where $w_m$ is a maximum or minimum in $w(a)$.

This is a completely general result, applying to all quintessence evolution.
However, it provides particularly interesting constraints on some of the models considered here.  For example, consider the quadratic potential given by
Eq. (\ref{Vquadratic}), with $V_0$ and $V_2$ related through the value of $K$
given in Eq. (\ref{quadK}).  For this case, we have
\begin{equation}
\frac{V^\prime}{V} = \frac{6(1-K^2)\phi}{8 + 3(1-K^2)\phi^2}.
\end{equation}
For $K^2 > 1$, corresponding to potentials with $V^{\prime \prime} < 0$, $(V^\prime/V)^2$ can be arbitrarily large.
However, for $V^{\prime\prime} > 0$
(i.e., $K^2 < 1)$, we see that $(V^\prime/V)^2$ takes on a maximum value at $\phi = \sqrt{8/3(1-K^2)}$, so that
\begin{equation}
\left(\frac{V^\prime}{V}\right)^2 < \frac{3}{8}(1-K^2).
\end{equation}
Then Eq. (\ref{wvalue}) gives
\begin{equation}
\label{wm}
1+w_m < \frac{1}{8}(1-K^2) \Omega_\phi.
\end{equation}
In the examples considered here, when the scalar field first freezes , we have $\Omega_\phi << 1$ and $(1-K^2) \sim O(1)$,
so Eq. (\ref{wm}) implies that $w$ is driven to a value nearly equal to $-1$ before thawing again.  This is indeed
the behavior we observe.
Then, since
$\Omega_\phi < \Omega_{\phi_0}$, Eq. (\ref{wm}) gives an upper bound on $w$ after the field begins to thaw, namely
(for $\Omega_{\phi 0} \sim 0.7$),
\begin{equation}
\label{limit}
1+w < 0.09 (1-K^2).
\end{equation}
Thus, although it might appear that one can obtain arbitrarily large values of $w_0$ by appropriate choices of $w_i$ and and $\phi_i$,
Eq. (\ref{limit}) shows that this is not the case for $K^2 < 1$, and we find that our numerical method does, indeed, fail to find a solution when
$w_0$ is increased above this upper bound.

\begin{figure}[!t]
\centerline{\epsfxsize=4truein\epsffile{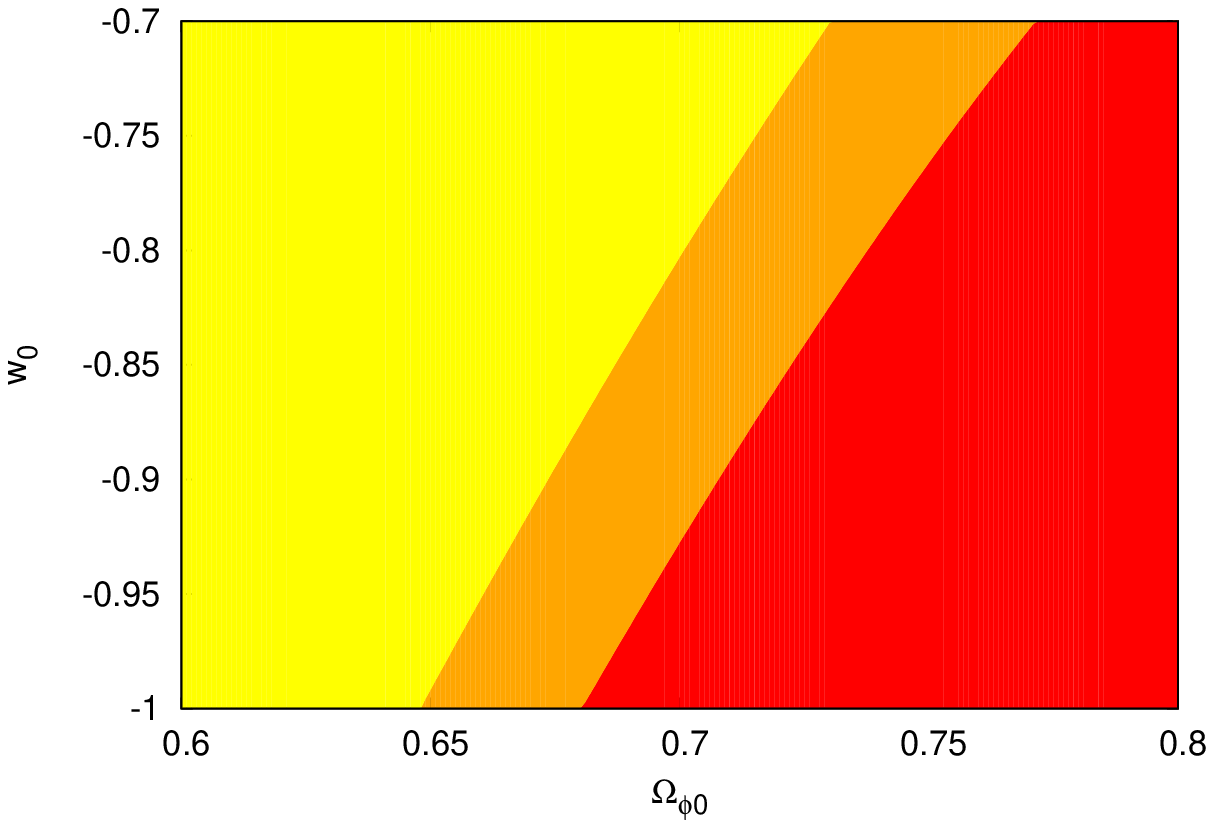}}
\caption{Likelihood plot from SNIa data in the  $w_0 - \Omega_{\phi 0}$ plane for
the quadratic potential for the case corresponding to the
($+$) solution with $K^2 = 10$ and $w_i = -0.8$ at scale factor $a_i = 0.2$.  Yellow
region is excluded at $2\sigma$, orange region is excluded at $1\sigma$, and
red region is not excluded at either confidence level.}
\end{figure}

\begin{figure}[!t]
\centerline{\epsfxsize=4truein\epsffile{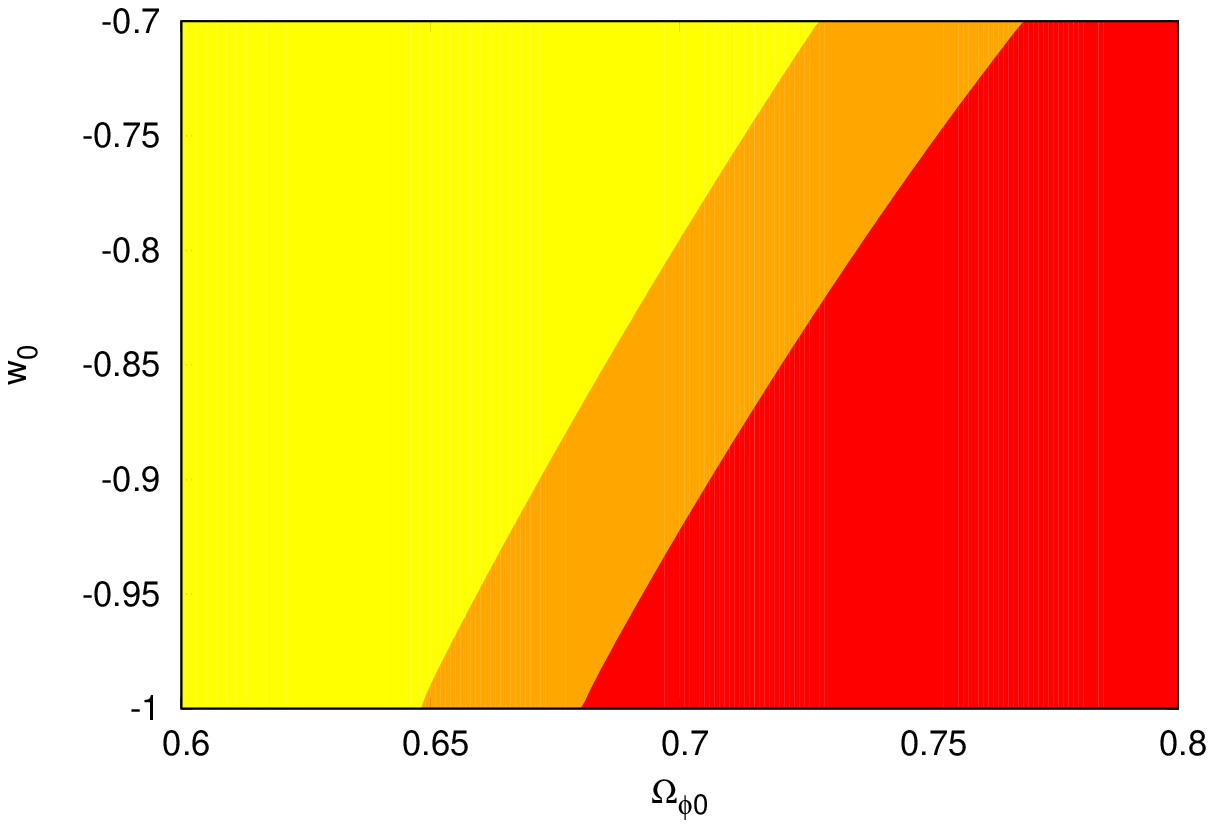}}
\caption{As Fig. 7 for the ($-$) solution with 
$K^2 = 10$.}
\end{figure}

\begin{figure}[!t]
\centerline{\epsfxsize=4truein\epsffile{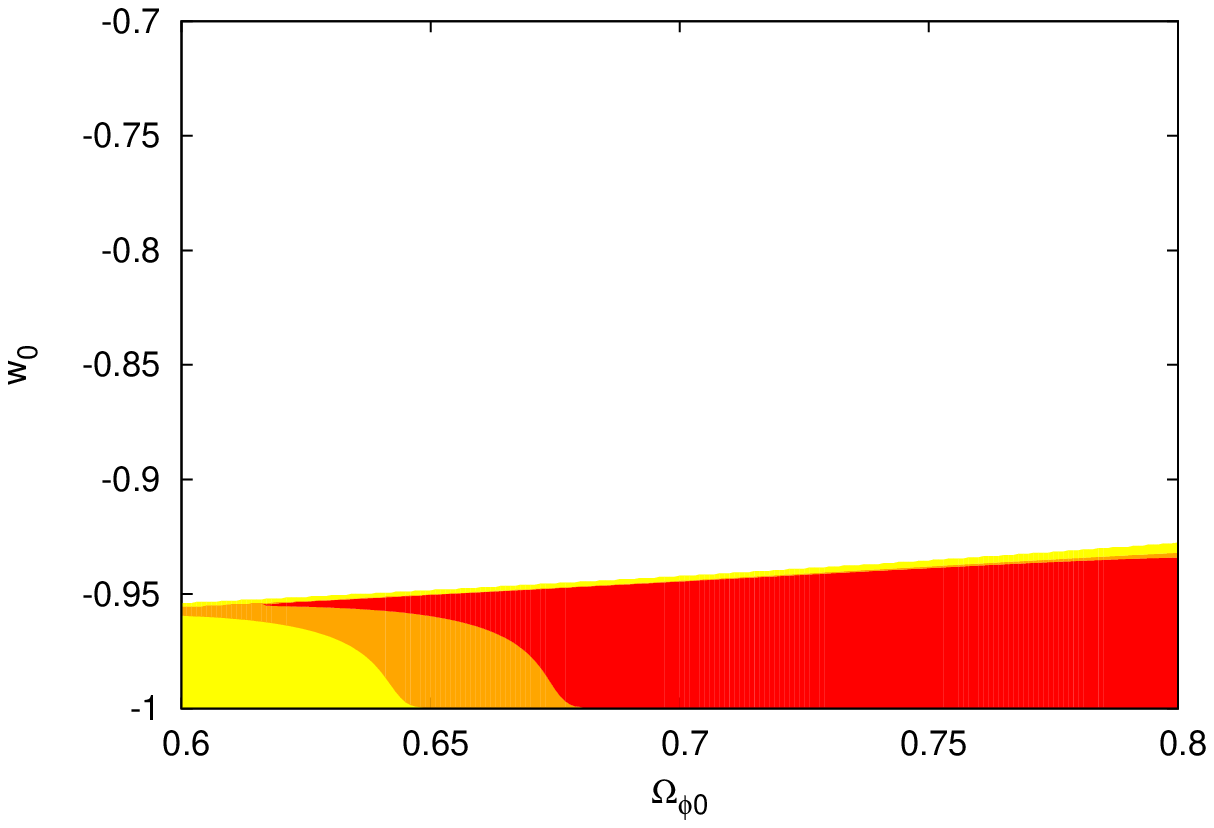}}
\caption{As Fig. 7 for the ($+$) solution with 
$K^2 = 0$.  Unshaded space represents a final value of $w_0$ that cannot
be attained for these parameters (see Sec. IV).}
\end{figure}

\begin{figure}[!t]
\centerline{\epsfxsize=4truein\epsffile{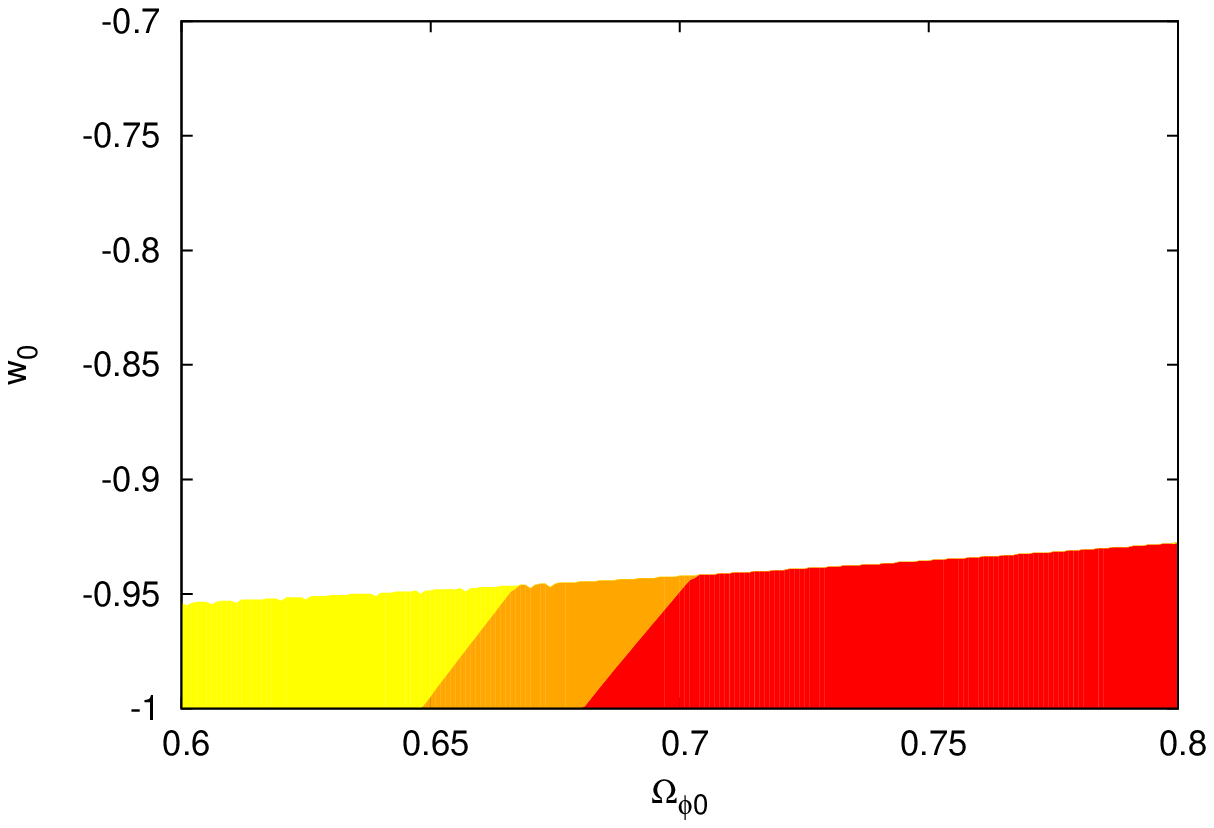}}
\caption{As Fig. 7 for the ($-$) solution with 
$K^2 = 0$. Unshaded space represents a final value of $w_0$ that cannot
be attained for these parameters (see Sec. IV).}
\end{figure}

\begin{figure}[!t]
\centerline{\epsfxsize=4truein\epsffile{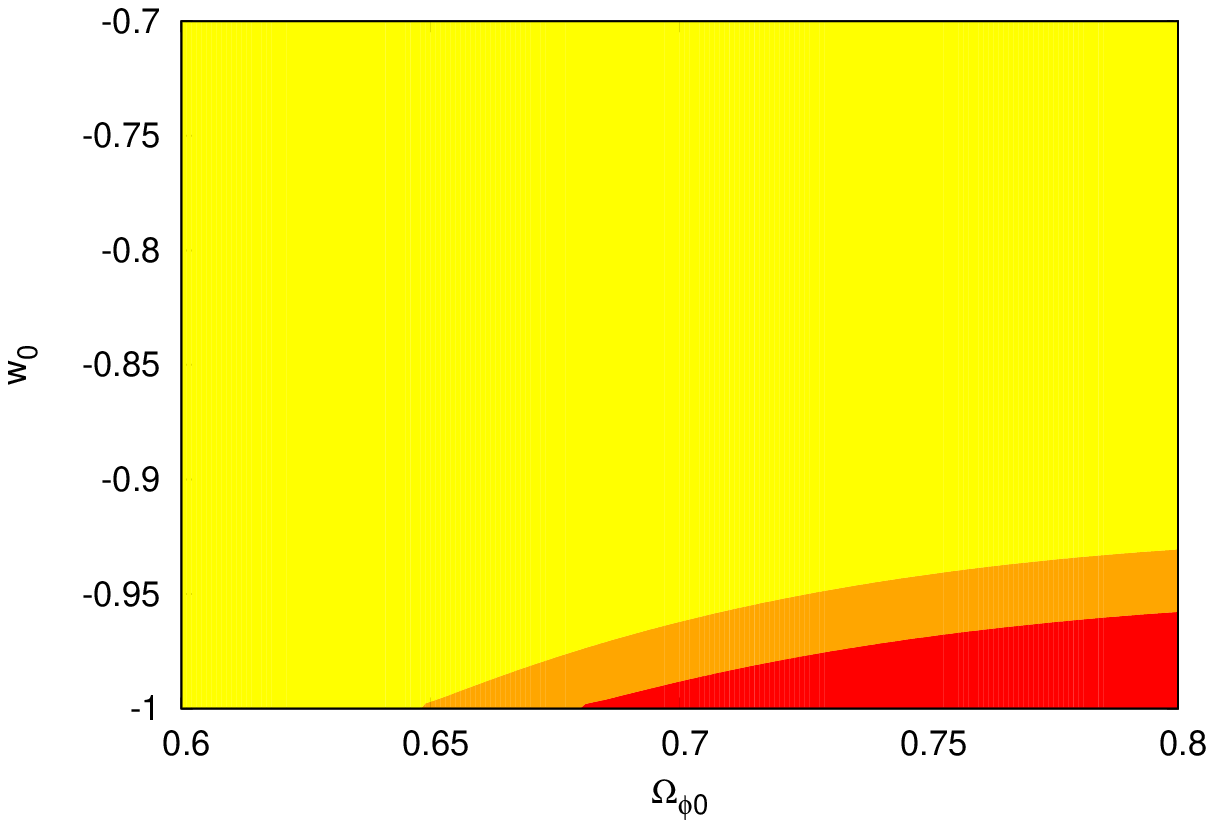}}
\caption{As Fig. 7 for the ($+$) solution with 
$K^2 = -10$.}
\end{figure}

\begin{figure}[!t]
\centerline{\epsfxsize=4truein\epsffile{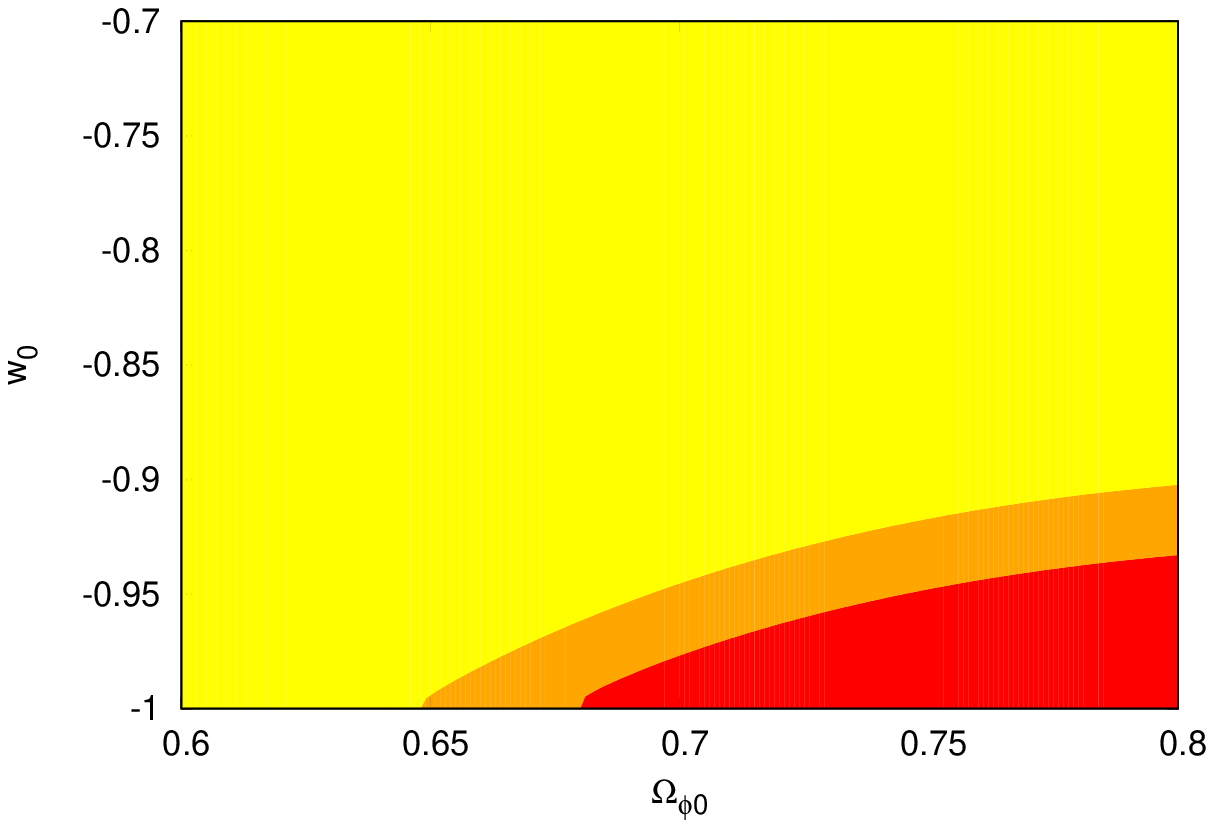}}
\caption{As Fig. 7 for the ($-$) solution with 
$K^2 = -10$.}
\end{figure}

\section{Comparison with Observations}

Since, as noted in Sec. III, our general expression (Eqs. \ref{wfinal}$-$\ref{gK}) provides a reasonable
fit to a variety of potentials, this expression can
be used in conjunction with observational data to constrain such potentials.
In order to produce such constraints, we performed a $\chi^2$-test of the SCP Union 2.1 dataset, presented in Ref. \cite{Suzuki}, with the numerically solved
quadratic potential.  The evolution of $\phi$ for a
given $\Omega_{\phi 0}$ and $K^2$ determines the Hubble parameter according to Eqs. (\ref{density}) and (\ref{H2}),
which can in turn be used to numerically solve for the luminosity distance,
\begin{equation}
d_L(z)=(1+z)\int_{0}^{z}\frac{dz'}{H(z')},
\end{equation}
as a function of redshift.  The distance modulus can be extracted via the relation
\begin{equation}
\mu=5\log_{10}(d_L)-5
\end{equation}
in order to compare our model to the SNe Ia data.

We have four free parameters, $\Omega_{\phi0}$, $w_0$, $K$, and $w_i$; further,
$w_i$
must itself be chosen at some initial scale factor $a_i$.  We have taken
$a_i = 0.2$ and scanned over a variety of values for $K$ and $w_i$ to derive
likelihoods in the $\Omega_{\phi0}$, $w_0$ plane.  In what
follows we present our results for
$w_i = -0.8$, and $K^2 = -10$, $0$, and $+10$.  Note that for each value of $w_i$,
there are two different values of $\dot \phi_i$, corresponding to the field
rolling initially in different directions in the potential.  These
correspond to the ($+$) and ($-$) solutions in Eq. (\ref{wfinal}).

The likelihood contours are presented in Figs. $(7)-(8)$ for $K^2 = 10$, in Figs.
$(9)-(10)$ for $K^2 = 0$ and in Figs. $(11)-(12)$ for $K^2 = -10$.  The first thing to note
is that the likelihoods are relatively insensitive to the value of $\dot \phi_i$, as shown
by the fact that the likelihood contours do not change much in going from the ($+$) to
the ($-$) solutions, at least for $|K^2| \gg 1$.  The potentials with $K^2 = 10$ (corresponding to
hilltop models) produce a much larger allowed parameter space than $K^2 = -10$
(corresponding to models with $V^{\prime\prime} > 0$).  But perhaps most interesting, in the models
with $K^2=0$, the biggest constraint on the parameter space comes not from the observational
data, but from the dynamics of the scalar field itself.  In the case, $w_0$ is automatically
constrained to be very close to $-1$ by the arguments presented in Sec. IV.

\section{Discussion}

Our analytic expression provides excellent agreement with the exact numerical evolution for all of the potentials we have examined, but
we find the best fit
for the PNGB potential, which is also the best-motivated of the models we have considered.  Similar
excellent agreement for this model, for the special case of $\dot \phi_i = 0$, was noted by Dutta and Scherrer
\cite{ds1}, but the reasons for this are not at all clear, since the analytic expression was designed to approximate the quadratic potential.
We have presented observational limits on these models, but perhaps the most interesting limit is our upper bound on the final value
of $w$ for models with sufficiently small curvature in the potential.

Our analytic expression for $w(a)$ represents a final generalization of the framework constructed in Refs. 
\cite{ScherrerSen1,ScherrerDutta1,ds1,ds3}.  Indeed, all of the results in Refs. \cite{ScherrerSen1,ScherrerDutta1,ds1,ds3}
are special cases of our Eqs. (\ref{wfinal})$-$(\ref{gK}).  Taking $1+w_i = 0$, our equations reduce to the
results in Refs. \cite{ds1,ds3}.  Taking the limit where $K \rightarrow 1$ (but $K \ne 1$), we obtain the results
in Ref. \cite{ScherrerDutta1}, and if we take both $1+w_i = 0$ and $K \rightarrow 1$, we regain the results of Ref. \cite{ScherrerSen1}.
Just as in Ref. \cite{ScherrerDutta1}, all of our solutions begin with an initial freezing evolution, but then at late times
they evolve as in Refs. \cite{ds1,ds3}.  One previous study that is {\it not} subsumed within the results presented here is the paper
by Chiba \cite{Chiba}, who does not assume evolution near a local potential maximum or minimum.  However, it would be possible
to generalize Chiba's results to the case of a nonzero $\dot \phi_i$, using the methods we have outlined here.

\section{Acknowledgments}
R.J.S. was supported in part by the Department of Energy
(DE-FG05-85ER40226).  We thank Dan Li for helpful comments on the manuscript.


\begin{thebibliography}{99}

\bibitem{Knop}
R.A. Knop, et al., Ap.J. {\bf 598}, 102 (2003).

\bibitem{Riess}
A.G. Riess, et al., Ap.J. {\bf 607}, 665 (2004).


\bibitem{union08}
  M.~Kowalski {\it et al.},
  %``Improved Cosmological Constraints from New, Old and Combined Supernova
  %Datasets,''
  Astrophys.\ J.\  {\bf 686}, 749 (2008).
%  [arXiv:0804.4142 [astro-ph]].
  %%CITATION = ASJOA,686,749;%%
  
\bibitem{perivol}
  L.~Perivolaropoulos and A.~Shafieloo,
  \prd {\bf 79}, 123502 (2009).
  %``Bright High z SnIa: A Challenge for LCDM?,''
%  arXiv:0811.2802 [astro-ph].
  %%CITATION = ARXIV:0811.2802;%%

\bibitem{hicken}
M.~Hicken {\it et al.},
\apj {\bf 700}, 1097 (2009).
  %``Improved Dark Energy Constraints from ~100 New CfA Supernova Type Ia Light
  %Curves,''
%  arXiv:0901.4804 [astro-ph.CO].
  %%CITATION = ARXIV:0901.4804;%%



%%%%%%CMB

\bibitem{Hinshaw}
G. Hinshaw, et al., Ap.J. Suppl. {\bf 208}, 19 (2013).
  
\bibitem{Ade}
P.A.R. Ade, et al., [arXiv:1303.5076].
  
  \bibitem{Wetterich}
  C. Wetterich,
  Nucl. Phys. B {\bf 302}, 668 (1988).
  
  \bibitem{RatraPeebles}
  B.~Ratra and P.~J.~E.~Peebles,
  %``Cosmological Consequences of a Rolling Homogeneous Scalar Field,''
  Phys.\ Rev.\  D {\bf 37}, 3406 (1988).
  %%CITATION = PHRVA,D37,3406;%%
  
  
%\cite{Caldwell:1997ii}
\bibitem{CaldwellDaveSteinhardt}
  R.~R.~Caldwell, R.~Dave and P.~J.~Steinhardt,
  %``Cosmological Imprint of an Energy Component with General
  %Equation-of-State,''
  Phys.\ Rev.\ Lett.\  {\bf 80}, 1582 (1998).
  %%CITATION = PRLTA,80,1582;%%
  
 %\cite{Liddle:1998xm}
\bibitem{LiddleScherrer}
  A.~R.~Liddle and R.~J.~Scherrer,
  %``A classification of scalar field potentials with cosmological scaling
  %solutions,''
  Phys.\ Rev.\  D {\bf 59}, 023509 (1999).
  %%CITATION = PHRVA,D59,023509;%%
  
\bibitem{SteinhardtWangZlatev}
  P.~J.~Steinhardt, L.~M.~Wang and I.~Zlatev,
  %``Cosmological Tracking Solutions,''
  Phys.\ Rev.\  D {\bf 59}, 123504 (1999).
  %%CITATION = PHRVA,D59,123504;%%
  
\bibitem{Copeland}
E.J. Copeland, M. Sami, and S. Tsujikawa, Int. J. Mod. Phys. D
{\bf 15}, 1753 (2006).

  
\bibitem{ScherrerSen1}
R.~J.~Scherrer and A.~A.~Sen,
%``Thawing quintessence with a nearly flat potential,''
Phys.\ Rev.\  D {\bf 77}, 083515 (2008)
%  [arXiv:0712.3450 [astro-ph]].
%%CITATION = PHRVA,D77,083515;%%


  
\bibitem{ds1}
S.~Dutta and R.~J.~Scherrer,
\prd {\bf 78}, 123525 (2008).
  %``Hilltop Quintessence,''
%  [arXiv:0809.4441 [astro-ph]].
  %%CITATION = PHRVA,D78,123525;%%
  
%\cite{Dutta:2009yb}
\bibitem{ds3}
  S.~Dutta, E.~N.~Saridakis and R.~J.~Scherrer,
  \prd {\bf 79}, 103005 (2009).
  %``Dark energy from a quintessence (phantom) field rolling near potential
  %minimum (maximum),''
%  arXiv:0903.3412 [astro-ph.CO].
  %%CITATION = ARXIV:0903.3412;%%

\bibitem{ScherrerDutta1}
S. Dutta and R.J. Scherrer,
Phys. Lett. B {\bf 704}, 265 (2011).

\bibitem{Gong}
Y. Gong, Phys. Lett. B {\bf 731}, 342 (2014).

\bibitem{CL}
R.R. Caldwell and E.V. Linder, \prl {\bf 95}, 141301 (2005).

\bibitem{Chiba}
T. Chiba, \prd {\bf 79}, 083517 (2009); erratum \prd {\bf 80}, 109902 (2009).

\bibitem{Huiyiing}
H.-Y. Chang and R.J. Scherrer, \prd {\bf 88}, 083003 (2013).

\bibitem{Frieman}
J.A. Frieman, C.T. Hill, A. Stebbins, and I. Waga,
\prl {\bf 75}, 2077 (1995).

\bibitem{Dutta}
K. Dutta and L. Sorbo,
\prd {\bf 75}, 063514 (2007).

\bibitem{Albrecht}
A. Abrahamse, A. Albrecht, M. Barnard, and B. Bozek,
\prd {\bf 77}, 103503 (2008).

\bibitem{LinderPNGB}
R. de Putter and E.V. Linder, 
JCAP {\bf 10}, 042 (2008).

\bibitem{Brax}
P. Brax and J. Martin, Phys. Lett. B {\bf 468}, 40 (1999).

\bibitem{Linder}
E.V. Linder, \prd {\bf 73}, 063010 (2006).

\bibitem{Suzuki}
N. Suzuki, {\it et al.}, Ap.J. {\bf 746} 85 (2012)

\end{thebibliography}
\end{document}